\documentclass[12pt,preprint]{aastex}

\begin{document}

\title{Collapse and Fragmentation of Magnetic Molecular Cloud Cores with 
the Enzo AMR MHD Code. II. Prolate and Oblate Cores}

\author{Alan P.~Boss and Sandra A.~Keiser}
\affil{Department of Terrestrial Magnetism, Carnegie Institution for
Science, 5241 Broad Branch Road, NW, Washington, DC 20015-1305}
\email{boss@dtm.ciw.edu}

\begin{abstract}

 We present the results of a large suite of three-dimensional (3D) models of the 
collapse of magnetic molecular cloud cores using the adaptive mesh refinement (AMR) 
code Enzo2.2 in the ideal magnetohydrodynamics (MHD) approximation. The cloud cores 
are initially either prolate or oblate, centrally condensed clouds with masses of 
1.73 or 2.73 $M_\odot$, respectively. The radial density profiles are Gaussian, 
with central densities 20 times higher than boundary densities. A barotropic equation 
of state is used to represent the transition from low density, isothermal phases, 
to high density, optically thick phases. The initial magnetic field strength 
ranges from 6.3 to 100 $\mu$G, corresponding to clouds that are 
strongly to marginally supercritical, respectively, in terms of the mass 
to magnetic flux ratio. The magnetic field is initially uniform and aligned 
with the clouds' rotation axes, with initial ratios of rotational to gravitational 
energy ranging from $10^{-4}$ to 0.1.  Two significantly different outcomes 
for collapse result: (1) formation of single protostars with spiral 
arms, and (2) fragmentation into multiple protostar systems. The transition 
between these two outcomes depends primarily on the initial magnetic field 
strength, with fragmentation occurring for mass to flux ratios greater than 
about 14 times the critical ratio for prolate clouds. Oblate clouds typically 
fragment into several times more clumps than prolate clouds. Multiple, rather 
than binary, system formation is the general rule in either case, 
suggesting that binary stars are primarily the result of the orbital dissolution 
of multiple protostar systems. 

\end{abstract}


\section{Introduction}

 While the Sun appears to be a single star, binary and multiple stars are 
commonplace. The comprehensive survey of all solar-type (F6 to K3) main-sequence 
stars within 25 pc by Raghavan et al. (2010) found $\sim$ 56\% of the primary stars
to have no confirmed companions, $\sim$ 33\% to have a binary companion, $\sim$ 
8\% to be in triple systems, and $\sim$ 1\% to be in even higher-order systems.
When counting the total number of stars involved, then, this means that at least
$\sim$ 2/3 of all stars are members of binary or multiple star systems. In fact, 
increasingly higher binary star frequencies are found in increasingly 
younger stellar and protostellar populations (e.g., Class O protostars; 
Chen et al. 2013), implying that essentially all stars could be born in multiple 
systems, a fraction of which eventually decay through orbital evolution and close 
encounters, leading to the ejection of single stars, and leaving behind stable binary 
and hierarchical multiple systems (Reipurth et al. 2014). The leading mechanism for 
the formation of binary and multiple stars is fragmentation (Reipurth et al. 2014), 
where a dense molecular cloud core undergoes dynamic collapse and rapidly 
breaks up into two or more protostellar objects (Boss \& Bodenheimer 1979).
Brown dwarf binaries with total system masses as small as $\sim 0.02 M_\odot$ 
have also been detected (Choi et al. 2013), implying that binary fragmentation 
continues to operate in this sub-stellar mass regime as well. 

 Observations of Zeeman splitting in molecular clouds have shown that magnetic 
fields can often be detected, with field strengths significant enough 
to affect the outcome of the fragmentation process (e.g., Girart et al. 2013). 
Crutcher (2012) concluded that while magnetic field strengths generally 
are more important than turbulence for supporting molecular clouds against
collapse, these fields are not strong enough to overcome gravity, at least
not for clouds with column densities higher than $N_H \sim 10^{21}$ cm$^{-2}$:
such relatively dense clouds appear to be magnetically supercritical, and
hence should be either contracting or susceptible to contraction. As a result,
3D MHD calculations of cloud collapse have become commonplace in recent
years (Reipurth et al. 2014), supplanting the 3D hydrodynamics (HD) that
had previously been considered state-of-the-art (e.g., Bonnell et al. 2007). 
Hennebelle et al. (2011) used an AMR MHD code to follow the collapse of a 
relatively massive, 100 $M_\odot$ cloud, finding that the initial magnetic 
field strength reduced the amount of fragmentation by as much as a factor 
of two, compared to a nonmagnetic cloud collapse. Clearly magnetic fields
need to be considered when assessing the efficiency of the fragmentation
mechanism for forming binary and multiple star systems.

 Observations of starless, pre-collapse molecular cloud cores have shown 
that dense cloud cores are centrally-condensed (e.g., Ward-Thompson, Motte, 
\& Andr\'e 1999; Schnee et al. 2010; Ruoskanen et al. 2011; Butler \& Tan 
2012), highly non-spherical, and often turbulent. Velocity gradients 
indicative of overall rotation in dense cloud cores lead to estimates 
of the ratio of rotational to gravitational energy ranging from 
$\sim 10^{-4}$ to $\sim 0.1$ (Caselli et al. 2002; Harsono et al. 2014).
Prolate and oblate shapes have been inferred from 
observations of suspected pre-collapse molecular cloud cores (e.g., 
Jones, Basu, \& Dubinski 2001; Curry \& Stahler 2001; Cai \& Taam 2010;
Gholipour \& Nejad-Asghar 2013; Lomax, Whitworth, \& Cartwright 2013). 

 While Boss \& Keiser (2013; hereafter BK13) used Enzo to study the collapse of
initially uniform density, rotating, spherical clouds, here we consider the 
collapse of more realistic, centrally-condensed, spheroidal molecular cloud cores, 
namely the initially prolate and oblate cloud cores considered by Boss (2009), 
who used a pseudo-MHD code to model magnetic field effects and ambipolar
diffusion (e.g., Boss 2005, 2007). We intend in part to see if these pseudo-MHD 
calculations produced results similar to those obtained with a true MHD 
code like Enzo. Machida, Inutsuka, \& Matsumoro (2014) used an MHD code to
study the effects of starting from either initially uniform density or 
centrally-condensed cloud cores, and found that the initial density profile 
had a significant effect on the formation of protostellar disks. Here we focus
on the effects of the initial magnetic field strength on fragmentation
in initially centrally-condensed cloud cores. We seek to return to the
questions of the extent to which the fragmentation mechanism occurs in 
dense molecular cloud cores when magnetic fields are included, and when 
it does occur, what this implies for the formation of single, binary, 
and multiple protostellar systems.
 
\section{Numerical Methods}

 While BK13 used the Enzo2.0 code, here we have employed the updated Enzo2.2 
version, which includes several code improvements as well as bug fixes (e.g., 
BK13 found a bug in an early version of Enzo2.0 dealing with the gas mean 
molecular weight, which was corrected in later versions). Enzo offers
several choices for the hydrodynamics scheme (Collins et al. 2010; 
Bryan et al. 2014), which are designed to conserve mass and 
energy. Following BK13, we used the Runge-Kutta third-order-based MUSCL 
(monotone upstream-centered schemes for conservation laws) algorithm 
(HydroMethod = 4), which is based on the Godunov (1959) shock-handling HD 
method, along with the Harten-Lax-van Leer (HLL) Riemann solver. Ideal MHD was 
performed using the Dedner et al. (2002) divergence cleaning method (Wang 
et al. 2008). The divergence of the magnetic field was monitored throughout 
the calculations, in order to ensure that this quantity, which should remain
zero for ideal MHD, remained small throughout the evolutions. More 
details about these and other numerical matters may be found in BK13.

 The standard MHD and HD models were initialized in Enzo2.2 on a 3D 
Cartesian top grid with 64 grid points in each direction; several models 
used 128 grid points in each direction for comparison testing. The
standard models had a maximum of 6 levels of refinement that could
be implemented as needed during the collapses, each level with a factor of 
2 refinement over the previous level. Hence the maximum possible effective 
grid resolution for the standard models was $2^6 = 64$ times higher than 
the initial top grid resolution of $64^3$, i.e., $4096^3$. Several 
comparison models explored the effects of changing the maximum number of
levels of refinement from 6 to as high as 10 and to as low as 2 levels.
Grid refinement was performed automatically whenever necessary to ensure 
that the Jeans length constraint (e.g., Truelove et al. 1997) was satisfied 
by a factor of 8 or more. Reflecting boundary conditions were applied on
each face of the top grid's cubic box, after tests with the other boundary
conditions possible with Enzo showed that reflecting boundary conditions
did the best job of conserving mass and angular momentum (to within
1\% and 6\%, respectively) during non-magnetic collapse to central 
densities of $\sim 10^{-10}$ g cm$^{-3}$. Changing the top grid resolution
from $32^3$ to $64^3$, on the other hand, had no effect on the conservation 
of total mass and angular momentum. The maximum number of Green's functions
used to calculate the gravitational potential was 10, with 10 iterations
performed on the gravitational potential at each step. The time step used was 0.3 
of the limiting Courant time step. The second-order-accurate InterpolationMethod = 1
option was adopted to maximize mass and angular momentum conservation.
The yt astrophysical analysis and visualization toolkit (Turk et al. 2011)
was used to analyze the results.

 Collapsing molecular cloud cores remain isothermal at $\sim$ 10 K 
until the central regions reach densities of $\sim 10^{-14}$ g cm$^{-3}$ 
(e.g., Boss 1986), at which point the centers become optically thick to infrared
radiation, and compressional heating begins to raise the clouds' temperatures. 
While radiative transfer is necessary thereafter to follow the subsequent evolution 
properly (e.g., Boss 1986; Boss et al. 2000), here we explore the nonisothermal
regime by using the ``barotropic'' equation employed by, e.g., Price \& Bate 
(2007), B\"urzle et al. (2011), BK13, and many others. The barotropic 
pressure $p$ depends on the gas density $\rho$ as $p = K \rho^\gamma$, where the
polytropic exponent $\gamma$ equals 1 for $\rho < 10^{-14}$ g cm$^{-3}$
and 7/5 for $\rho > 10^{-14}$ g cm$^{-3}$. The factor $K = c_s^2$, where 
$c_s$ is the isothermal sound speed (0.2 km s$^{-1}$), below the critical 
density of $\rho_c = 10^{-14}$ g cm$^{-3}$, and $K = c_s^2 \rho_c^{-2/5}$
for $\rho > \rho_c$. This results in continuity of the pressure 
across the critical density. The barotropic approximation is a
variation of the ``adiabatic'' approximation for the nonisothermal regime 
introduced by Boss (1981), who noted that such calculations could be scaled
to clouds of arbitrary mass. The adiabatic approximation assumes
$p \propto \rho^\gamma$ with $\gamma = 7/5$, and since the Boss (1981) models
all started with very high densities, $\rho \sim 10^{-12}$ g cm$^{-3}$,
these adiabatic approximation models were effectively the same as 
barotropic approximation models. Future calculations will include radiative
transfer effects in the flux-limited diffusion approximation, which has
been implemented in Enzo (Reynolds et al. 2009), but which will increase
the computational burden substantially, compared to the barotropic
approximation. Radiative transfer was included in 3D HD ORION AMR
calculations by Krumholz et al. (2007) and Offner et. al (2009),
in 3D MHD RAMSES AMR calculations by Commercon et al. (2010), in 3D 
HD SPH calculations by Whitehouse \& Bate (2006) and Bate (2009, 2011, 2012),
and in 3D MHD SPH calculations by Price \& Bate (2009), all of whom
found that radiative transfer had important effects not captured by
the barotropic approximation.
 
 Rather than attempt to follow the nonisothermal collapse phase through 
to the formation of the first and second protostellar cores (e.g., 
Bate, Tricco, \& Price 2014), in some models we employ sink particles to 
represent the highest density clumps that form during protostellar collapse
and fragmentation. Sink particles are the moving versions of the 
stationary sink cells first introduced by Boss \& Black (1982) and
used to study the accretion of infalling gas by protostars. Sink particles 
have been developed for use in Enzo by Wang et al. (2010) and used by Zhao 
\& Li (2013). Sink particles have also been developed for use in smoothed 
particle hydrodynamics (SPH) calculations (Bate, Bonnell, \& Price 1995;
Hubber, Walch, \& Whitworth 2013), in the Athena code (e.g., Gong \& Ostriker 
2013), in GADGET-2 (e.g., B\"urzle et al. 2011; Klapp et al. 2014), and elsewhere. 
Recently Machida, Inutsuka, \& Matsumoto (2014) found that sink cells need to be 
less than 1 AU in size, and to have threshold densities greater than 
$n \sim 10^{13}$ cm$^{-3}$, in order to properly study the formation of 
disks during the main accretion phase of protostellar collapse.

 We used the Enzo sink particle coding described in some detail by
Wang et al. (2010). Sink particles are created in grid cells that have
already been refined to the maximum extent permitted by the specified
number of levels of grid refinement, but where the gas density still 
exceeds that consistent with the Jeans length criterion for avoiding
spurious fragmentation (Truelove et al. 1997; Boss et al. 2000). The sink 
particles thereafter accrete gas from their host cells at the modified 
Bondi-Hoyle accretion rate proposed by Ruffert (1994). Two parameters
control the conditions under which sink particles can be merged together:
the merging mass (SinkMergeMass) and the merging distance 
(SinkMergeDistance). The former of these two parameters is used to
divide the sink particles into either large or small particles. Particles
with less mass than SinkMergeMass are first subjected to being combined with
any large particles that are located within the SinkMergeDistance. Any surviving
small particles after this first step are then merged with any other small
particles within the SinkMergeDistance. The merging process is performed
in such a way as to ensure conservation of mass and linear momentum. Note
that the magnetic field is not accreted by the sink particles, which
can be considered as a crude means of depicting decoupling of the magnetic 
field from high density gas (Zhao \& Li 2013). Wang et al. (2010) and
Zhao \& Li (2013) discuss the extent to which AMR code results depend on the
choices made for the SinkMergeMass and SinkMergeDistance parameters; we have 
performed a similar investigation, with relatively large changes in these
two sink particle parameters (see Section 4.4).

 Ambipolar diffusion has been incorporated into the FLASH AMR code by
Duffin \& Pudritz (2008), the ZEUS 3D code (Mac Low et al. 1995), and
into other codes as well (e.g., Choi et al. 2009). Ambipolar diffusion
was included in early versions of Enzo, but it is not supported in the
most recent versions (Bryan et al. 2014). Collins et al. (2011) argued
that ambipolar diffusion was unimportant for their Enzo calculations of the
collapse of cloud cores because the initial magnetic field strengths
were assumed to be relatively weak (i.e., $B_i \sim 1 \mu$G for
$\rho_i \sim 10^{-21}$ g cm$^{-3}$). Boss (1997) derived an approximation
for ambipolar diffusion in a 3D pseudo-MHD code, based on true MHD
collapse models, and used this formalism in subsequent papers in the
series (e.g., Boss 2005, 2007, 2009). While one of our goals is to
compare our Enzo results with those of Boss (2009), we do not attempt
to include ambipolar diffusion in the present models, for several reasons. 
First, current versions of Enzo do not support ambipolar diffusion. Second,
the approximation derived by Boss (1997) was essentially one of lowering
the magnetic field strength in a linear manner until dynamic collapse
began, after which collapse occurs so rapidly that the field is essentially
frozen-in to the gas, i.e., ideal MHD dominates. Third, Crutcher (2012) 
has concluded on observational grounds that ambipolar diffusion is not 
important, but that magnetic reconnection is more likely to be important for
solving the magnetic flux problem of star formation (Lazarian, Esquivel, 
\& Crutcher 2012). Magnetic reconnection is effective on large scales in 
turbulent clouds (e.g., Santos-Lima, de Gouveia Dal Pino, \& Lazarian 2013).
We leave for future work the consideration of initially turbulent clouds, 
which are likely to experience significant magnetic reconnection.

\section{Initial Conditions}

 We study the collapse of initially prolate and oblate molecular cloud
cores, the same initial conditions studied by Boss (2009). Boss (2009) 
used a 3D HD code with radiative transfer in the Eddington approximation,
but only a pseudo-MHD handling of magnetic fields. Here we employ a true
MHD code, Enzo2.2, but resort to a simple barotropic equation of state
to represent the effects of radiative transfer in optically thick regions
on the gas pressure. 

 As in Boss (2009), the initial cloud cores are centrally condensed, with 
Gaussian radial density profiles (Boss 1997) of the form

\begin{equation}
\rho_i(x,y,z) = \rho_o \ exp\biggl(
- \biggl( {x \over r_a} \biggr)^2
- \biggl( {y \over r_b} \biggr)^2
- \biggl( {z \over r_c} \biggr)^2
\biggr),
\end{equation}

\noindent 
where $\rho_0 = 2.0 \times 10^{-18}$ g cm$^{-3}$ is the initial central 
density. The initial central density is roughly a factor of 20 times
higher than that at the edge of the cloud for all of the models. The 
clouds are initially either prolate or oblate, with axis ratios of 2 to 1. 
The prolate clouds have $r_a = 1.16 R$ and $r_b = r_c = 0.580 R$, where $R$ is 
the cloud radius, while the oblate clouds have $r_a = r_b = 1.16 R$ and 
$r_c = 0.580 R$. The cloud radius is $R = 1.0 \times 10^{17}$ cm =
0.032 pc; the top level AMR grid is a cube $2.0 \times 10^{17}$ cm on a 
side. Random numbers in the range [0,1] are used
to add noise to these initial density distributions by multiplying
$\rho_i$ from the above equation by the factor $[1 + 0.1 \ ran(x,y,z)]$.
This form of density perturbation was used, instead of a velocity
perturbation, in order to make the initial conditions as close as possible 
to those of Boss (2009), who used this same density perturbation with no
velocity perturbation. The cloud masses are 1.73 $M_\odot$ and 2.73 $M_\odot$, 
respectively, for the prolate and oblate clouds, slightly higher
than for the corresponding clouds in Boss (2009), 1.5 $M_\odot$ and 2.1 $M_\odot$, 
respectively, because of the cubical volume of the Enzo computational grid
compared to the spherical volume of the Boss (2009) spherical coordinate code
grid. With an initial temperature of 10 K, the initial ratio of thermal 
to gravitational energy for the Boss (2009) models is $\alpha_i = 0.39$ for 
the prolate clouds and 0.30 for the oblate clouds; such clouds are unstable
to collapse in the absence of rotation or magnetic fields.

 The initial magnetic field strengths assumed for the cloud cores range 
from 6.3 $\mu$G to 100 $\mu$G. Estimates of the line-of-sight component of the 
magnetic field in dense cloud cores with column densities appropriate for 
these models ($\sim 5 \times 10^{22}$ cm$^{-2}$) are on the order of
$B_{los} \sim 10 \mu$G to $\sim 200 \mu$G (Figure 7 in Crutcher 2012).
The critical mass $M_\Phi$ is given by $\Phi/(2 \pi G^{1/2})$, where $\Phi$ is 
the magnetic flux (Nakano \& Nakamura 1978; Crutcher 2012) and $G$ is the 
gravitational constant. Magnetically subcritical clouds have masses less than 
$M_\Phi$, while magnetically supercritical clouds have masses greater than 
$M_\Phi$ and should be able to contract and collapse. The critical mass to 
flux ratio is then defined by $\mu_c = M_\Phi/\Phi = 1/(2 \pi G^{1/2})$.
For the Gaussian profile prolate clouds studied here, the initial mass to flux
ratios range from $\sim$ 28 to 1.8 $\mu_c$ for the initial field strengths of 6.3 
to 100 $\mu$G, respectively, i.e., from clouds that are strongly to marginally 
supercritical, respectively. The initial mass to flux ratios are a factor
of $\sim 2.73/1.73 = 1.58$ times higher for the oblate clouds, i.e., ranging from
44.9 to 11.4 $\mu_c$ for 6.3 to 25 $\mu$G, respectively. Boss (2009) started
all models with a magnetic field strength of 200 $\mu$G, resulting in 
initially magnetically stable clouds with subcritical mass to flux ratios.
The approximate inclusion of ambipolar diffusion allowed the Boss (2009)
clouds to collapse eventually.

 Solid body rotation is assumed, with the angular velocity about 
the $\hat z$ axis taken to be $\Omega_i$ = $10^{-14}$, 
$3.2 \times 10^{-14}$, $10^{-13}$, or $3.2 \times 10^{-13}$
rad s$^{-1}$. These choices of $\Omega_i$ result in initial ratios of 
rotational to gravitational energy ($\beta_i$) ranging from $\sim 10^{-4}$ 
to $\sim 0.1$, for both the prolate and oblate clouds. Solid body rotation
is a reasonable approximation to continue to employ for the initial
conditions for dense cloud cores, as indications are that turbulence is dominant
only on larger scales than on the scale of individual dense cloud cores, where 
the velocity field is more likely to be ordered (Klapp et al. 2014). The 
orthogonal alignment of bipolar outflows with the position angles of protostellar
binary systems similarly implies that initial turbulence does not dominate the
fragmentation mechanism (Tobin et al. 2013). SPH calculations of the collapse
of turbulent molecular cloud cores lead to misalignments of the disks
formed with the orbital planes of the binary or multiple star systems
(Tsukamoto \& Machida 2013). 

 The models with non-zero magnetic fields all began with the same initial 
orientation: initially straight magnetic fields that were aligned with the 
$\hat z$ axis (i.e., parallel to the cloud's rotation axis). Krumholz, Crutcher, 
\& Hull (2013) found that magnetic field lines and rotational axes are 
randomly aligned in cloud cores. However, Chapman et al. (2013) found a 
positive correlation between bipolar outflow directions and core magnetic 
field directions, with the former presumably being indicative of the protostar's 
rotation axis. Li, Krasnopolsky, \& Shang (2013) studied the effects of
magnetic field misalignments with rotation axes, as did Price \& Bate (2007)
and BK13. Large-scale magnetic fields do not appear to be correlated with the 
orientations of spheroidal cloud cores, implying they do not shape
the cores (Frau et al. 2012). The Boss (2009) pseudo-MHD models assumed
the magnetic field to be aligned with the rotation axis, and used an
approximate treatment (Boss 2007) of magnetic tension forces to model
the effects of magnetic braking. BK13's true MHD calculations
found that the magnetic field alignment with respect to the cloud rotation
axis could change the critical initial field strength leading to collapse
and fragmentation by a factor of about two.

 Finally, we note that BK13 found that the angular momentum loss due to 
magnetic fields initially aligned with the rotation axis was small (only a
few percent, comparable to that lost due to non-conservation of
angular momentum by Enzo during non-magnetic collapse) unless the
initial mass to flux ratio was less than the critical value, i.e.,
unless the clouds were initially magnetically sub-critical, which 
is not the case for any of the present set of models.

\section{Results}

\subsection{Resolution Test Cases}

 We begin with a discussion of several test models with variations in 
the parameters that control the spatial resolution of the Enzo code, a
critical factor for fragmentation models (e.g., Truelove et al. 1997;
Boss et al. 2000). The intention is to settle upon a reasonable set of
standard Enzo parameters that can be used for the bulk of the MHD models.
BK13 found that a standard spatial resolution for the top grid of $32^3$
and a maximum of 6 levels of refinement was sufficient to demonstrate
the same outcome for the standard isothermal test case (Boss \& Bodenheimer 1979)
that had been found by previous high spatial resolution HD calculations,
namely a runaway thin bar, and obtained a similar result with a top grid
of $64^3$. Here we use $64^3$ as our standard top grid resolution, though
one model did explore a $128^3$ top grid (model BNPP2R2 in Table 1). We also
varied the maximum number of levels of refinement (MLR) from MLR = 2 to 4 
to 6 to 10 for variations on the collapse of several of the non-magnetic, 
prolate cloud models (PP2R through PP2E in Table 1), all of which resulted 
in collapse and fragmentation into multiple systems when MLR = 6 was used.
With MLR = 4, model PP2R-4 collapsed and formed a multiple system, but
with MLR = 2, model PP2R-2 collapsed and formed a single system, as the fragments
formed merged together early in the collapse. On the other hand, with MLR = 10,
models PP2A-10, PP2C-10, and PP2E-10 all collapsed to form thin bars with 
such high densities that the AMR routines generated a number of
sub-grid points large enough to exceed the memory available on the Carnegie Xenia
Cluster nodes and halt the computations. This occurred so early in the
collapses ($\sim 1.4 t_{ff}$, as defined below) that the bar wind-up and fragmentation 
phase (see below) could not be reached (at $\sim 1.6 t_{ff}$). Given these results,
a choice of MLR = 6 seemed to be best for achieving reasonably high resolution
as well as computational efficacy. Finally, with MLR = 4, five models were
run with variations in the choice of other Enzo mesh refinement parameters,
namely the MinimumMassForRefinement and the CellFlaggingMethod (e.g., using 
the baryon mass density for refinement), to ensure that these parameters
produced the correct outcome whether the Jeans length criterion was employed 
or not. All the models presented below used a standard safety factor of 8 in
the Jeans length constraint.

\subsection{Prolate Cloud Core Models}

 Table 1 lists the initial conditions and basic results for the MHD models 
starting with prolate cloud cores. The models span the applicable range of
initial cloud rotation rates and increasingly magnetically supercritical clouds.
The results column describes the density configurations (generally in the
midplane, $\hat z = 0$) at the final time computed $t_f$, expressed in terms 
of the free fall time at the initial central density, 
$t_{ff} = (3 \pi / 32 G \rho_o)^{1/2} = 4.72 \times 10^4$ yr,
where $\rho_o = 2.0 \times 10^{-18}$ g cm$^{-3}$. The final time was chosen to
be when the maximum density was achieved, typically $\sim 10^{-11}$ g cm$^{-3}$
to $\sim 10^{-10}$ g cm$^{-3}$. Many models were evolved considerably
farther in time for curiosity's sake, into a realm where the evolution of the 
highest density regions was probably not being accurately depicted, and hence
those later phases are not reported here.

 Figure 1 shows the typical evolution of a fragmenting cloud, in this case the
non-magnetic model PP2A. Starting from the initial prolate cloud, seeded with
random density perturbations (Figure 1(a)), the cloud collapses on the free-fall
time scale to form a thin bar (Figure 1(b)), as occurs in the isothermal collapse of 
clouds with initial $m = 2$ density perturbations (e.g., Boss \& Bodenheimer 1979, 
BK13). However, once densities rise above $10^{-14}$ g cm$^{-3}$, the nonisothermal
evolution phase of collapse begins, and the rapidly rising thermal pressure stops
collapse at the center of the cloud. As a result of this bounce, combined with
the ongoing accretion of higher angular momentum gas, the bar begins to wind
up and form a rotationally flattened protostellar disk (Figure 1(c)), which 
continues to gain mass until it becomes gravitationally unstable and fragments 
into a multiple protostellar system (Figure 1(d)). The clumps seen in Figure 1(d)
have masses ranging from 0.0051 to 0.13 $M_\odot$, a typical range of clump
masses for the models listed in Table 1 at the final times. These clumps
are the first protostellar cores, composed primarily of molecular hydrogen, 
with maximum temperatures of about 300 K.

 Figure 2 depicts the evolution of model BMPP2A, which is identical to model
PP2A, except for having a magnetic field with an initial mass to flux ratio
of 7.2 (Table 1). BMPP2A also collapses to form a high density bar (Figure 2(a)),
which proceeds to wind up and form a protostellar disk (Figure 2(b)). However, in
this case, the added effects of the magnetic field lead to the formation of
the transient protostellar ring (off-axis density maximum) seen in Figure 2(c).
Two density maxima form in the ring, but the magnetic field effects are sufficiently
strong to prohibit the robust fragmentation seen in Figure 1(d) for model PP2A,
and instead a single central protostar forms (Figure 2(d)) with a central
temperature of 300 K. The spiral arms evident in Figure 2(d), in combination
with magnetic torques, are sufficient to transport angular momentum outward,
allowing the central protostar to gain mass while avoiding fragmentation into
a multiple system. The disk begins to expand slightly as a result of the outward
angular momentum transport and the continued infall of higher angular momentum
portions of the initial cloud core. Figure 3 illustrates the expanded size
of the disk in model BMPP2A (green regions) compared to model PP2A, as well as 
the much thicker disk aspect ratio, and the filamentary structures associated
with ideal MHD effects.

  Figure 4 shows the results for model BNPP2A, identical to models PP2A and
BMPP2A except for having a mass to flux ratio of 14.4, twice that of BMPP2A.
Model BNPP2A collapses to form a bar that begins to wind up and form a
disk (Figure 4(a)), but in this case, the reduced magnetic effects are
unable to prevent the formation of a disk whose outer regions rapidly fragment 
into a multiple system surrounding a central protostar (Figure 4(b)).
This multiple system undergoes a chaotic evolution, with clumps forming
and merging (Figure 4(c)) as the maximum density continues to increase
(Figure 4(d)). At the final time, three distinct clumps remain in the form
of an inner binary and a hierarchical triple system, with clump masses of
0.034, 0.086, and 0.20 $M_\odot$, though other lower mass clumps persist as well 
(Figure 4(d)). The densest clump in the central binary has a maximum temperature 
close to 500 K. 

 A set of four models with a top grid of $128^3$ was run (BNPP2R2,
BNPP2A2, BNPP2C2, and BNPP2E2 in Table 1), which were otherwise identical 
to the corresponding models with a $64^3$ top grid (BNPP2R, BNPP2A, BNPP2C, 
and BNPP2E in Table 1). The results were qualitatively similar for both
sets of models, with the exception of models BNPP2A2 and BNPP2A, where
the higher resolution model (BNPP2A2) did not fragment by the final time
calculated. This was a result of the initially higher resolution allowing a
higher central density at an earlier time than model BNPP2R, so that the
central clump in the protostellar disk was able to dominate the disk, gaining
mass as its spiral arms transported angular momentum outward, and preventing
the formation of competing clumps. However, the higher initial rotation rate of
model BNPP2R2 was able to overcome this effect, and the central protostar
and disk eventually fragmented into a binary protostar system with masses
of 0.040 and 0.019 $M_\odot$.

 Table 1 also notes that two models formed single objects with disks 
whose midplanes were oriented out of the expected $z = 0$ plane, namely model
BPP2E ($x = 0$) and model BLPP2E ($y = 0$). Both of these models started with the
smallest initial rotation rates and the strongest initial magnetic fields, 
and hence were the models where magnetic field effects were expected to 
dominate the collapse. Evidently in these models the magnetic fields
were able to flip the orientation of the disks orbiting the central protostars.
This effect may be related to an effect found in the 3D MHD Enzo calculations by 
Zhao et al. (2011), who found that magnetic field-matter decoupling occurred in their 
sink particle calculations, leading to a magnetic pressure build-up in a low density
region close to the central sink particle, through which the field escaped.
While sink particles were not employed here in models BPP2E and BLPP2E, in both
models a similar density and magnetic field configuration developed as
seen by Zhao et al. (2011): low density bubbles filled with strong magnetic
fields, adjoining the highest density, central region. This interesting
effect is worthy of further investigation (see also Krasnopolsky et al. 2012),
though it does not appear to have a major influence on whether a given cloud 
can fragment or not, which is the focus of the present work. In a similar
context, the models with initially high values of $B_z = 50$ and 100 $\mu$G
in Table 1 all formed single protostars with small-scale, wide outflows along
the $\hat z$ axis, even model BPP2E with the flipped disk. Such outflows have 
also been found in many previous MHD collapse models (e.g., Tomisaka 2002; 
Machida et al. 2006, 2008; B\"urzle et al. 2011).

\subsection{Oblate Cloud Core Models}

 Table 2 lists the initial conditions and basic results for the MHD models 
starting with oblate cloud cores, most of which collapsed and fragmented.
Figure 5 displays the evolution of a representative oblate cloud core,
model BNPO2A, which collapsed to form a dense oblate disk (Figure 5(b))
before undergoing rotational and magnetic rebound (Figure 5(c)) and 
fragmenting into a small cluster of clumps, with masses ranging from 0.038 
to 0.53 $M_\odot$, the latter with a central temperature of nearly 500 K.

 Whereas Table 1 showed that a mass to flux
ratio of about 14 times the critical ratio separates prolate cloud cores that 
collapse and fragment from those that collapse to form single protostars, 
Table 2 shows that for initially oblate cloud cores, the initial mass to flux
ratio only has to be less than about 11 in order for fragmentation to be avoided. 
Figure 6 compares the results for four oblate clouds that are identical
except for their initial magnetic to flux ratios, namely models BMPO2R,
BNPO2R, BOPO2R, and PO2R, with $M/\Phi$ ratios of 11.4, 22.8, 44.9, and
$\infty$, respectively, of the critical value. Figure 6 clearly shows
the progressively stronger fragmentation efficiency as the initial $M/\Phi$
ratio increases, showing that for an initial $M/\Phi$ less than $\sim 11$,
fragmentation is as impeded as it is for the prolate cloud cores
with $M/\Phi$ less than $\sim 14$. All things being equal, an oblate cloud is
more likely to be able to fragment during collapse than a prolate cloud, given
the greater amount of cloud mass at larger initial cloud radius, and thus
higher specific angular momentum, so one expects that the critical $M/\Phi$
ratio should be somewhat lower for oblate clouds than for prolate clouds,
as seems to be the case for these models, though given the coarseness
of the changes in $M/\Phi$ ratio (i.e., by factors of two), this is not
definitive. It is also evident that the oblate clouds tend to produce
more clumps when fragmentation occurs than prolate clouds, e.g., oblate 
models PO2R, PO2A, PO2C, and PO2E produced a total of 31 clumps, whereas
the corresponding prolate models PP2R, PP2A, PP2C, and PP2E produced a
total of 11 clumps at the final time.

\subsection{Sink Particle Test Cases}

 Sink particles offer the promise of allowing 3D HD or MHD calculations
to be continued much farther in time than would otherwise be the case,
given the constraints placed on times steps by the Courant condition
and on spatial resolution by the need to resolve the Jeans length. 
Here we briefly explore the reliability of Enzo's sink particles to
depict both fragmentation and the evolution of the sink particles
that form, by running a set of models (Table 3) with variations 
in the merging mass (SinkMergeMass) and the merging distance 
(SinkMergeDistance) for sink particles, both described in Section 2.
The models are otherwise identical, except for changes by factors
of 10, in each direction, for these parameters, compared to the
standard values used in model BOPO2RSP (Table 3) and in the sink
particle models to follow.

 Figure 7 shows the outcome of four of the models from Table 3, at
comparable final times. It is evident from Figure 7 that while the details
vary for each of these models, as expected given the chaotic orbital
evolution of closely-packed, gravitationally interacting bodies,
nevertheless even with significant changes in these two key parameters,
the sink particle technique robustly predicts fragmentation into
a large number of protostellar clumps. In fact, Table 3 shows that the
final number of sink particles only varies from 5 to 8, and the total amount 
of mass in the sink particles is always within the range of 1.48 to 1.63
$M_\odot$, showing reasonably good overall agreement. The masses of the
sink particles in these five models are also similar, with less than
a factor of 2 spread in the maximum particle mass, and a somewhat larger
variation in the minimum particle mass. These models indicate that the
results of sink particle representations of fragmenting clouds are fairly
robust, at least to within factors of $\sim$ 2.

\subsection{Prolate Cloud Core Models with Sink Particles}

 Table 4 lists the initial conditions and basic results for the MHD models 
starting with prolate cloud cores with sink particles. Figure 8 shows
the results for four models that are identical except for having
variations in the initial rotation rates. The fastest rotating models,
BOPP2RS and BOPP2AS, each fragment into 3 sink particles (Figure 8(a,b)),
while the two slower rotating models (BOPP2CS and BOPP2ES) collapse to form 
single sink particles (Figure 8(c,d)). Comparing to the corresponding models 
without sink particles in Table 1 (BOPP2R, BOPP2A, BOPP2C, and BOPP2E),
all of which fragmented into multiple clumps, the sink particle models
show a reduced tendency to form long-lasting fragments. This is largely
due to the fact that the sink particle models can be advanced considerably
farther in time than the models without sink particles: Table 4 shows that
the former reach final times of 2.60 to 4.30 $t_{ff}$, compared to only
1.85 to 2.59 $t_{ff}$ for the latter. Once fragmentation has occurred,
continued orbital evolution is likely to only reduce the total number
of clumps or sink particles, following orbital mergers or ejections.
In addition, by their very definition,
the sink particles are designed to accrete nearby clumps and lower-mass
particles, in order to speed the computation, which will result in
fewer sink particles (e.g., Table 3 shows that the total number of sink particles
is reduced when the SinkMergeDistance is increased to 333 AU, and the opposite 
when SinkMergeDistance is decreased to 3.3 AU).

\subsection{Oblate Cloud Core Models with Sink Particles}

 Finally, Table 5 lists the initial conditions and basic results for the MHD models 
starting with oblate cloud cores with sink particles. Figure 9 shows
the results for four models that are identical except for having
variations in the initial rotation rates. The fastest rotating models,
BNPO2RS and BNPO2AS, fragment into either 6 or 3 sink particles, respectively
(Figure 9(a,b)), while the two slower rotating models (BNPO2CS and BNPO2ES) 
collapse to form single sink particles (Figure 9(c,d)). These outcomes
are quite similar to those found for the initial prolate clouds with
sink particles, shown in Figure 8. Once again, compared to the corresponding
oblate cloud models without sink particles (Table 2), the inclusion of
sink particles results in single protostars for the slower rotating clouds,
as is to be expected, given the advanced times achieved with the sink
particle technique, and the susceptibility of the slower rotating clouds
to having their clumps be merged into a single protostar. Table 5 shows
that by the final times, the sink particles are typically able to accrete 
more than half of the total oblate cloud mass of 2.73 $M_\odot$.

 Figure 10 compares the time evolution of the clump masses in the oblate
cloud model BNPO2R with the sink particle masses in the corresponding
model BNPO2RS. While model BNPO2R shows a wide variation in clump
masses once fragmentation begins after 1 $t_{ff}$, the sink particles
in model BNPO2RS show a more monotonic increase in the minimum and maximum sink
particle masses. Overall, the clump masses and sink particle masses
are in basic agreement, at least at the level of factors of two.

\section{Discussion}

 Two distinct outcomes characterize these magnetically supercritical clouds: 
collapse and formation of a single protostar with significant spiral arms, 
or else collapse and fragmentation into binary or multiple protostar systems, 
with multiple spiral arms. For prolate clouds, the critical magnetic field
strength is about 12 $\mu$G, or a mass to flux ratio of about 14 times the
critical ratio (Table 1), whereas for oblate clouds, the critical magnetic field
strength is about 25 $\mu$G, or a mass to flux ratio of about 11 times the
critical ratio (Table 2). These results are comparable to the results found
by BK13, who studied the barotropic fragmentation of initially uniform density,
spherical, solar-mass clouds with magnetic fields initially parallel to the cloud 
rotation axis, finding that a mass to flux ratio of about 13 times the critical 
ratio (corresponding to a magnetic field strength of about 70 $\mu$G) 
separated clouds that collapsed and fragmented from those that collapsed to 
form single protostars (see Table 4 in BK13). These results suggest that
the mass to flux ratio of a given initial cloud is a more robust predictor
of the outcome of dynamic collapse than the initial magnetic field strength.
Joos et al. (2013) also studied the collapse and fragmentation
of initially centrally condensed cloud cores with an ideal MHD code, finding
that fragmentation occurred for mass to flux ratios greater than 5 times
the critical ratio, roughly a factor of 2 times smaller than found here.
However, their initial cloud masses were 5 $M_\odot$, and contained
significant turbulent velocity fields, which served to misalign the
magnetic field direction compared to the overall cloud rotation axis.
Joos et al. (2013) noted that fragmentation did not occur for this mass
to flux ratio when the cloud mass was 1 $M_\odot$, showing that their
results depended strongly on the initial cloud mass. As a result of these
factors, the results obtained here for centrally condensed clouds with 
masses of 1.73 and 2.73 $M_\odot$ appear to be in reasonable agreement
with the results of Joos et al. (2013).

 We now compare to the results of Boss (2009), who used a pseudo-MHD code
with radiative transfer to study the 3D collapse of prolate and oblate
spheroids identical to those studied here (i.e., compare Tables 1 and 2
of Boss (2009) with the present Tables 1 and 2, respectively). The Boss (2009)
models all started with an initial magnetic field strength $B_o = 200 \mu$G,
i.e., with magnetically sub-critical clouds, but used a prescription
for ambipolar diffusion to reduce the reference magnetic field strength
over a nominal ambipolar diffusion time scale of $\tau_{AD} = 10 t_{ff}$.
As a result, by the time that collapse occurred, the nominal field strength 
$B_o$ had dropped to $\sim 100 \mu$G for the 20 to 1 density ratio clouds 
also studied here, equivalent to starting with an initial mass to flux
ratio of about 2 times the critical value, i.e., super-critical. Boss (2009) 
found mixed results: the prolate clouds fragmented into quadruple, binary, 
or binary-bar configurations as the initial rotation rate dropped from
$10^{-13}$ to $10^{-14}$ rad s$^{-1}$, whereas the oblate clouds
over that same span of rotation rates either did not collapse significantly,
or collapsed to form a ring with no evidence for fragmentation.
Given the differences between the Boss (2009) models and those presented
here, a more precise comparison does not appear to be possible, except
to note in general terms that lowering the initial rotation rate tends
to stifle fragmentation, and that magnetically super-critical clouds
with mass to flux ratios significantly greater than the critical value
are necessary for collapse and fragmentation into multiple protostar
systems.

 The present calculations rely on the barotropic approximation, which
is a convenient assumption for searching a large parameter space of
3D MHD AMR models, but which ultimately needs to be replaced by a proper 
treatment of radiative transfer, as demonstrated by Boss et al. (2000).
Radiative transfer has been included in cloud collapse models calculated
with both AMR (Krumholz et al. 2007; Offner et al. 2009; Commercon et al. 2010) 
and SPH codes (Whitehouse and Bate 2006; Price \& Bate 2009; Bate 2009, 
2011, 2012). The AMR calculations generally found that including radiative
transfer led to less fragmentation in circumprimary disks than in models with 
the barotropic approximation, as the radiative heating from the newly formed
primary protostar heated the disk enough to forestall gravitational 
instability. The SPH calculations found a similar trend, with fewer
brown dwarfs formed when radiative transfer was included (Bate 2009), a 
result in better agreement with observations. However, Bate (2011) found that
radiative transfer resulted in protostellar disks with lifetimes as much 
as three times longer, the most massive of which might be susceptible 
to fragmentation. 

 Bate (2012) studied the effects of radiation from newly formed stars 
during the collapse and early evolution of a 500 $M_\odot$ cloud that 
was initially highly Jeans unstable. Bate (2012) found a similar result 
to Bate (2009) for the collapse of a non-magnetic cloud with a mass ten times 
lower: fewer brown dwarfs. Bate's (2012) distribution of semi-major axes for the 
binary and multiple systems that resulted had a maximum for separations in the
range of 1 to 100 AU, similar to the observed distribution, which
peaks at $\sim$ 30 AU (Raghavan et al. 2010). However, Bate's (2012)
calculations resulted in more triple and higher order systems than
binary systems, compared to the statistics for solar-type stars,
where binary systems dominate by a factor of three or 
more (Raghavan et al. 2010). This could imply that magnetic fields,
which were not included in Bate (2012), play an essential
role in discouraging fragmentation enough to produce the correct
distribution and statistics for solar-type stars. Alternatively,
this discrepancy could to be due to the loss of some of these multiple
systems through their expected chaotic post-formational orbital evolution 
in the middle of a protostellar cluster. In any event, the present 
set of models serves to help define the extent to which magnetic fields can
stifle fragmentation, depending on the precise initial conditions
of a molecular cloud core.

\section{Conclusions}
  
 These 3D ideal MHD models have shown that for the case of rotating, centrally-condensed, 
barotropic, magnetic cloud cores, mass to flux ratios of about 14 times the 
critical ratio separate clouds that collapse to form single protostars and 
those that collapse and fragment into multiple protostar systems. For roughly
solar-mass clouds, this critical value corresponds to a magnetic field
strength of about 10 $\mu$G, which falls in the midst of the observed
values of $B_{los}$ and their lower limits (Figure 7 in Crutcher 2012).
Hence it appears that magnetic molecular cloud cores span the range
of $B_{los}$ values needed to produce both single and multiple protostar
systems. The calculations produce clumps with masses in the range of
$\sim 0.01$ to 0.5 $M_\odot$, clumps which will continue to accrete mass
and interact gravitationally with each other. It can be expected that
the multiple systems will undergo dramatic subsequent orbital evolution,
through a combination of mergers and ejections following close encounters,
resulting ultimately in a small cluster of stable hierarchical multiple
protostars, binary systems, and single protostars. Such evolution appears
to be necessary in order produce the binary and multiple star statistics
that hold for the solar-type stars in the solar neighborhood 
(Raghavan et al. 2010).

 We note that the use of sink particles resulted in fewer fragments at the final 
times, in large part because they allow the calculations to be followed significantly
farther in time, allowing clumps to merge together. However, given the approximate
treatment of protostellar physics in the sink particles (e.g., the assumption
that they do not accrete magnetic flux or angular momentum, compared to the clumps, 
which do contain significant magnetic flux and angular momentum), it is unclear if 
the sink particle results are to be trusted over the models without sink particles. 
In both cases, the inclusion of physical effects that have not been 
included in the present set of models (e.g., magnetic reconnection, non-ideal MHD, 
and radiative transport) is needed to decide whether 
either treatment gives a realistic assessment of the protostellar fragmentation 
process. Evidently there is much remaining to be studied with 3D AMR MHD codes.

\acknowledgments

 We thank Bo Reipurth for discussions about binary and multiple star
systems, the referee for a number of improvements to the manuscript,
and Nathan Goldbaum, John Wise, and Bo Zhao for advice about sink
particles in Enzo. The computations were performed using the Enzo code 
developed by the Laboratory for Computational Astrophysics at the 
University of California San Diego (http://lca.ucsd.edu). The 
calculations were performed on the Carnegie Xenia Cluster, the purchase 
of which was partially supported by the National Science Foundation (NSF) 
under grant MRI-9976645. This work was partially supported by NSF grant
AST-1006305. We thank Michael Acierno and Ben Pandit for their 
invaluable assistance with the Xenia cluster.

\clearpage
\begin{deluxetable}{cccccccc}
\tablecaption{Initial Conditions for Prolate Cloud Core Models with Varied Initial
Angular Velocities ($\Omega_i$), Ratios of Rotational to Gravitational
Energy ($\beta_i$), Magnetic Field Strengths ($B_z$), Mass to Flux 
Ratios ($M/\Phi$, in Units of the Critical Ratio), and Results (at the Final 
Time $t_f$, in Free Fall Units) in the Midplane ($x, y,$ or $z = 0$) Listed.
\label{tbl-1}}
\tablehead{\colhead{Model} & 
\colhead{$\Omega_i$ (rad s$^{-1}$) \quad} &
\colhead{$\beta_i$ \quad} &
\colhead{$B_z$ ($\mu$G) \quad} &
\colhead{$M/\Phi$ \quad} &
\colhead{$t_f/t_{ff}$ \quad} & 
\colhead{Result \quad } &
\colhead{Midplane \quad} } 
\startdata
BPP2R   & $3.2 \times 10^{-13}$ & $10^{-1}$  & 100.0 &  1.8     & 2.67 & single & z \\ 
BPP2A   & $1.0 \times 10^{-13}$ & $10^{-2}$  & 100.0 &  1.8     & 1.68 & single & z \\ 
BPP2C   & $3.2 \times 10^{-14}$ & $10^{-3}$  & 100.0 &  1.8     & 1.71 & single & z \\ 
BPP2E   & $1.0 \times 10^{-14}$ & $10^{-4}$  & 100.0 &  1.8     & 1.64 & single & x \\ 

\hline

BLPP2R  & $3.2 \times 10^{-13}$ & $10^{-1}$  &  50.0 &  3.6     & 1.86 & single & z \\ 
BLPP2A  & $1.0 \times 10^{-13}$ & $10^{-2}$  &  50.0 &  3.6     & 1.52 & single & z \\ 
BLPP2C  & $3.2 \times 10^{-14}$ & $10^{-3}$  &  50.0 &  3.6     & 1.52 & single & z \\ 
BLPP2E  & $1.0 \times 10^{-14}$ & $10^{-4}$  &  50.0 &  3.6     & 1.47 & single & y \\ 

\hline

BMPP2R  & $3.2 \times 10^{-13}$ & $10^{-1}$  &  25.0 &  7.2     & 2.34 & single & z \\ 
BMPP2A  & $1.0 \times 10^{-13}$ & $10^{-2}$  &  25.0 &  7.2     & 1.62 & single & z \\ 
BMPP2C  & $3.2 \times 10^{-14}$ & $10^{-3}$  &  25.0 &  7.2     & 1.61 & single & z \\ 
BMPP2E  & $1.0 \times 10^{-14}$ & $10^{-4}$  &  25.0 &  7.2     & 1.60 & single & z \\ 

\hline

BNPP2R  & $3.2 \times 10^{-13}$ & $10^{-1}$  &  12.5 &  14.4    & 2.76 & multiple & z \\ 
BNPP2A  & $1.0 \times 10^{-13}$ & $10^{-2}$  &  12.5 &  14.4    & 2.17 & multiple & z \\ 
BNPP2C  & $3.2 \times 10^{-14}$ & $10^{-3}$  &  12.5 &  14.4    & 1.92 & single & z \\ 
BNPP2E  & $1.0 \times 10^{-14}$ & $10^{-4}$  &  12.5 &  14.4    & 1.92 & single & z \\ 

\hline

BNPP2R2 & $3.2 \times 10^{-13}$ & $10^{-1}$  &  12.5 &  14.4    & 3.23 & binary & z \\ 
BNPP2A2 & $1.0 \times 10^{-13}$ & $10^{-2}$  &  12.5 &  14.4    & 1.74 & single & z \\ 
BNPP2C2 & $3.2 \times 10^{-14}$ & $10^{-3}$  &  12.5 &  14.4    & 1.64 & single & z \\ 
BNPP2E2 & $1.0 \times 10^{-14}$ & $10^{-4}$  &  12.5 &  14.4    & 1.61 & single & z \\ 

\hline

BOPP2R  & $3.2 \times 10^{-13}$ & $10^{-1}$  &  6.3  &  28.4    & 2.59  & multiple & z\\ 
BOPP2A  & $1.0 \times 10^{-13}$ & $10^{-2}$  &  6.3  &  28.4    & 1.91  & multiple & z \\ 
BOPP2C  & $3.2 \times 10^{-14}$ & $10^{-3}$  &  6.3  &  28.4    & 1.85  & multiple & z \\ 
BOPP2E  & $1.0 \times 10^{-14}$ & $10^{-4}$  &  6.3  &  28.4    & 1.88  & multiple & z \\ 

\hline

PP2R    & $3.2 \times 10^{-13}$ & $10^{-1}$  &  0.0  & $\infty$ & 2.11  & multiple & z \\ 
PP2A    & $1.0 \times 10^{-13}$ & $10^{-2}$  &  0.0  & $\infty$ & 1.68  & multiple & z \\ 
PP2C    & $3.2 \times 10^{-14}$ & $10^{-3}$  &  0.0  & $\infty$ & 1.66  & multiple & z \\ 
PP2E    & $1.0 \times 10^{-14}$ & $10^{-4}$  &  0.0  & $\infty$ & 1.66  & multiple & z \\ 
\enddata
\end{deluxetable}

\clearpage

\begin{deluxetable}{ccccccc}
\tablecaption{Initial Conditions for Oblate Cloud Core Models with Varied Initial
Angular Velocities ($\Omega_i$), Ratios of Rotational to Gravitational
Energy ($\beta_i$), Magnetic Field Strengths ($B_z$), Mass to Flux 
Ratios ($M/\Phi$, in Units of the Critical Ratio), and Results (at the Final 
Time $t_f$, in Free Fall Units).
\label{tbl-2}}
\tablehead{\colhead{Model} & 
\colhead{\quad $\Omega_i$ (rad s$^{-1}$) \quad} &
\colhead{\quad $\beta_i$ \quad} &
\colhead{\quad $B_z$ ($\mu$G) \quad} &
\colhead{\quad $M/\Phi$ \quad} &
\colhead{\quad $t_f/t_{ff}$ \quad} & 
\colhead{\quad Result \quad } } 
\startdata
BMPO2R  & $3.2 \times 10^{-13}$ & $10^{-1}$  &  25.0 &  11.4    & 2.03 & multiple  \\ 
BMPO2A  & $1.0 \times 10^{-13}$ & $10^{-2}$  &  25.0 &  11.4    & 1.39 & single    \\ 
BMPO2C  & $3.2 \times 10^{-14}$ & $10^{-3}$  &  25.0 &  11.4    & 1.36 & single    \\ 
BMPO2E  & $1.0 \times 10^{-14}$ & $10^{-4}$  &  25.0 &  11.4    & 1.62 & single    \\ 

\hline

BNPO2R  & $3.2 \times 10^{-13}$ & $10^{-1}$  &  12.5 &  22.8    & 2.38 & multiple  \\ 
BNPO2A  & $1.0 \times 10^{-13}$ & $10^{-2}$  &  12.5 &  22.8    & 1.51 & multiple  \\ 
BNPO2C  & $3.2 \times 10^{-14}$ & $10^{-3}$  &  12.5 &  22.8    & 1.50 & multiple  \\ 
BNPO2E  & $1.0 \times 10^{-14}$ & $10^{-4}$  &  12.5 &  22.8    & 1.55 & multiple  \\ 

\hline

BOPO2R  & $3.2 \times 10^{-13}$ & $10^{-1}$  &  6.3  &  44.9    & 2.15 & multiple  \\ 
BOPO2A  & $1.0 \times 10^{-13}$ & $10^{-2}$  &  6.3  &  44.9    & 1.37 & multiple  \\ 
BOPO2C  & $3.2 \times 10^{-14}$ & $10^{-3}$  &  6.3  &  44.9    & 1.50 & multiple  \\ 
BOPO2E  & $1.0 \times 10^{-14}$ & $10^{-4}$  &  6.3  &  44.9    & 1.42 & multiple  \\ 

\hline

PO2R    & $3.2 \times 10^{-13}$ & $10^{-1}$  &  0.0  & $\infty$ & 2.31 & multiple  \\ 
PO2A    & $1.0 \times 10^{-13}$ & $10^{-2}$  &  0.0  & $\infty$ & 1.48 & multiple  \\ 
PO2C    & $3.2 \times 10^{-14}$ & $10^{-3}$  &  0.0  & $\infty$ & 1.69 & multiple  \\ 
PO2E    & $1.0 \times 10^{-14}$ & $10^{-4}$  &  0.0  & $\infty$ & 1.62 & multiple  \\ 
\enddata
\end{deluxetable}

\clearpage

\begin{deluxetable}{cccccccc}
\tablecaption{Oblate Cloud Core Models with Sink Particles, Calculated
With Variations in the SinkMergeMass (SMM) and the SinkMergeDistance (SMD),
and Listing the Total, Maximum, Minimum Mass, and Total Number of Sink Particles 
($M_{SPtot}, M_{SPmax}, and M_{SPmin}$ in $M_\odot$, $N_{SPtot}$, respectively) 
at the Final Time.
\label{tbl-3}}
\tablehead{\colhead{Model} & 
\colhead{SMM $(M_\odot)$} &
\colhead{SMD (AU)} &
\colhead{$t_f/t_{ff}$} & 
\colhead{$M_{SPtot}$} &
\colhead{$M_{SPmax}$} &
\colhead{$M_{SPmin}$} &
\colhead{$N_{SPtot}$} } 
\startdata

BOPO2RSP &  0.01  & 33  & 2.69 & 1.63 & 0.387 & 0.115  &  6 \\ 
BOPO2RSQ &  0.01  & 333 & 2.61 & 1.56 & 0.422 & 0.176  &  5 \\ 
BOPO2RSR &  0.1   & 33  & 2.54 & 1.48 & 0.292 & 0.109  &  7 \\  
BOPO2RSS &  0.01  & 3.3 & 2.57 & 1.68 & 0.281 & 0.145  &  8 \\
BOPO2RST &  0.001 & 33  & 2.52 & 1.53 & 0.460 & 0.0522 &  6 \\ 

\enddata
\end{deluxetable}

\clearpage

\begin{deluxetable}{cccccccc}
\tablecaption{Prolate Cloud Core Models with Sink Particles, as in Table 1, with the 
Total Mass in Sink Particles, $M_{SPtot}$.
\label{tbl-4}}
\tablehead{\colhead{Model} & 
\colhead{$\Omega_i$ (rad s$^{-1}$) \quad} &
\colhead{$\beta_i$ \quad} &
\colhead{$B_z$ ($\mu$G) \quad} &
\colhead{$M/\Phi$ \quad} &
\colhead{$t_f/t_{ff}$ \quad} & 
\colhead{Result \quad } &
\colhead{$M_{SPtot} (M_\odot)$ \quad} } 
\startdata
BNPP2RS & $3.2 \times 10^{-13}$ & $10^{-1}$  &  12.5 &  14.4    & 5.82 & multiple & 0.470 \\ 
BNPP2AS & $1.0 \times 10^{-13}$ & $10^{-2}$  &  12.5 &  14.4    & 2.83 & multiple & 1.14 \\ 
BNPP2CS & $3.2 \times 10^{-14}$ & $10^{-3}$  &  12.5 &  14.4    & 3.63 & single   & 1.33 \\ 
BNPP2ES & $1.0 \times 10^{-14}$ & $10^{-4}$  &  12.5 &  14.4    & 2.59 & single   & 1.10 \\ 

\hline

BOPP2RS & $3.2 \times 10^{-13}$ & $10^{-1}$  &  6.3  &  28.4    & 4.30 & multiple & 0.627 \\ 
BOPP2AS & $1.0 \times 10^{-13}$ & $10^{-2}$  &  6.3  &  28.4    & 3.11 & multiple & 1.28 \\ 
BOPP2CS & $3.2 \times 10^{-14}$ & $10^{-3}$  &  6.3  &  28.4    & 2.60 & single   & 1.14 \\ 
BOPP2ES & $1.0 \times 10^{-14}$ & $10^{-4}$  &  6.3  &  28.4    & 3.63 & single   & 1.41 \\ 

\hline

PP2RS   & $3.2 \times 10^{-13}$ & $10^{-1}$  &  0.0  & $\infty$ & 3.84 & multiple & 0.791 \\ 
PP2AS   & $1.0 \times 10^{-13}$ & $10^{-2}$  &  0.0  & $\infty$ & 2.89 & single   & 0.818 \\ 
PP2CS   & $3.2 \times 10^{-14}$ & $10^{-3}$  &  0.0  & $\infty$ & 2.79 & single   & 1.36 \\ 
PP2ES   & $1.0 \times 10^{-14}$ & $10^{-4}$  &  0.0  & $\infty$ & 2.90 & single   & 1.40 \\ 
\enddata
\end{deluxetable}

\clearpage

\begin{deluxetable}{cccccccc}
\tablecaption{Oblate Cloud Core Models with Sink Particles, as in Table 2, with the 
Total Mass in Sink Particles, $M_{SPtot}$.
\label{tbl-5}}
\tablehead{\colhead{Model} & 
\colhead{$\Omega_i$ (rad s$^{-1}$) \quad} &
\colhead{$\beta_i$ \quad} &
\colhead{$B_z$ ($\mu$G) \quad} &
\colhead{$M/\Phi$ \quad} &
\colhead{$t_f/t_{ff}$ \quad} & 
\colhead{Result \quad } &
\colhead{$M_{SPtot} (M_\odot)$ \quad} } 
\startdata
BMPO2RS & $3.2 \times 10^{-13}$ & $10^{-1}$  &  25.0 &  11.4    & 3.48 & multiple & 1.42 \\ 
BMPO2AS & $1.0 \times 10^{-13}$ & $10^{-2}$  &  25.0 &  11.4    & 1.43 & single   & 0.309 \\ 
BMPO2CS & $3.2 \times 10^{-14}$ & $10^{-3}$  &  25.0 &  11.4    & 1.50 & single   & 0.955 \\ 
BMPO2ES & $1.0 \times 10^{-14}$ & $10^{-4}$  &  25.0 &  11.4    & 1.83 & single   & 1.74 \\ 

\hline

BNPO2RS & $3.2 \times 10^{-13}$ & $10^{-1}$  &  12.5 &  22.8    & 2.72 & multiple & 1.63 \\ 
BNPO2AS & $1.0 \times 10^{-13}$ & $10^{-2}$  &  12.5 &  22.8    & 1.84 & multiple & 1.65 \\ 
BNPO2CS & $3.2 \times 10^{-14}$ & $10^{-3}$  &  12.5 &  22.8    & 1.47 & single   & 0.887 \\ 
BNPO2ES & $1.0 \times 10^{-14}$ & $10^{-4}$  &  12.5 &  22.8    & 1.85 & single   & 1.70 \\ 

\hline

BOPO2RS & $3.2 \times 10^{-13}$ & $10^{-1}$  &  6.3  &  44.9    & 2.58 & multiple & 1.55 \\ 
BOPO2AS & $1.0 \times 10^{-13}$ & $10^{-2}$  &  6.3  &  44.9    & 2.00 & binary   & 1.83 \\ 
BOPO2CS & $3.2 \times 10^{-14}$ & $10^{-3}$  &  6.3  &  44.9    & 1.62 & single   & 1.39 \\ 
BOPO2ES & $1.0 \times 10^{-14}$ & $10^{-4}$  &  6.3  &  44.9    & 2.59 & single   & 2.21 \\ 

\hline

PO2RS   & $3.2 \times 10^{-13}$ & $10^{-1}$  &  0.0  & $\infty$ & 1.72 & multiple & 0.748 \\ 
PO2AS   & $1.0 \times 10^{-13}$ & $10^{-2}$  &  0.0  & $\infty$ & 1.77 & binary   & 1.78 \\ 
PO2CS   & $3.2 \times 10^{-14}$ & $10^{-3}$  &  0.0  & $\infty$ & 2.96 & single   & 2.64 \\ 
PO2ES   & $1.0 \times 10^{-14}$ & $10^{-4}$  &  0.0  & $\infty$ & 1.72 & single   & 1.83 \\ 
\enddata
\end{deluxetable}

\clearpage

\begin{figure}
\vspace{-2.0in}
\plotone{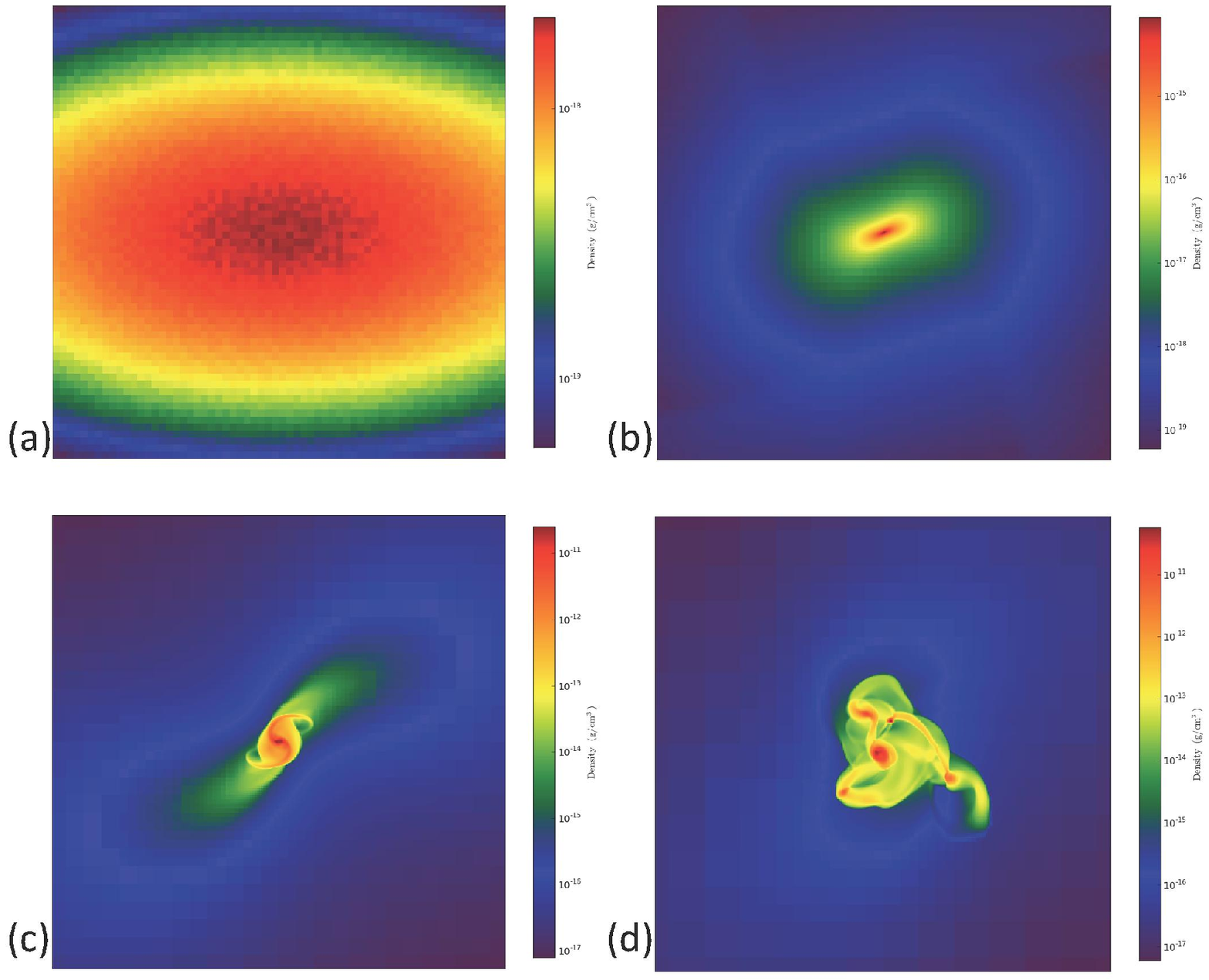}
\caption{Evolution of the midplane ($z = 0$) density distribution for model PP2A, shown 
at four times: (a) 0.0, (b) 1.40 $t_{ff}$, (c) 1.51 $t_{ff}$, and (d) 1.68 $t_{ff}$. 
Region shown is $2.0 \times 10^{17}$ cm across in (a) and (b) and 
$2.0 \times 10^{16}$ cm across in (c) and (d).}
\end{figure}

\clearpage

\begin{figure}
\vspace{-2.0in}
\plotone{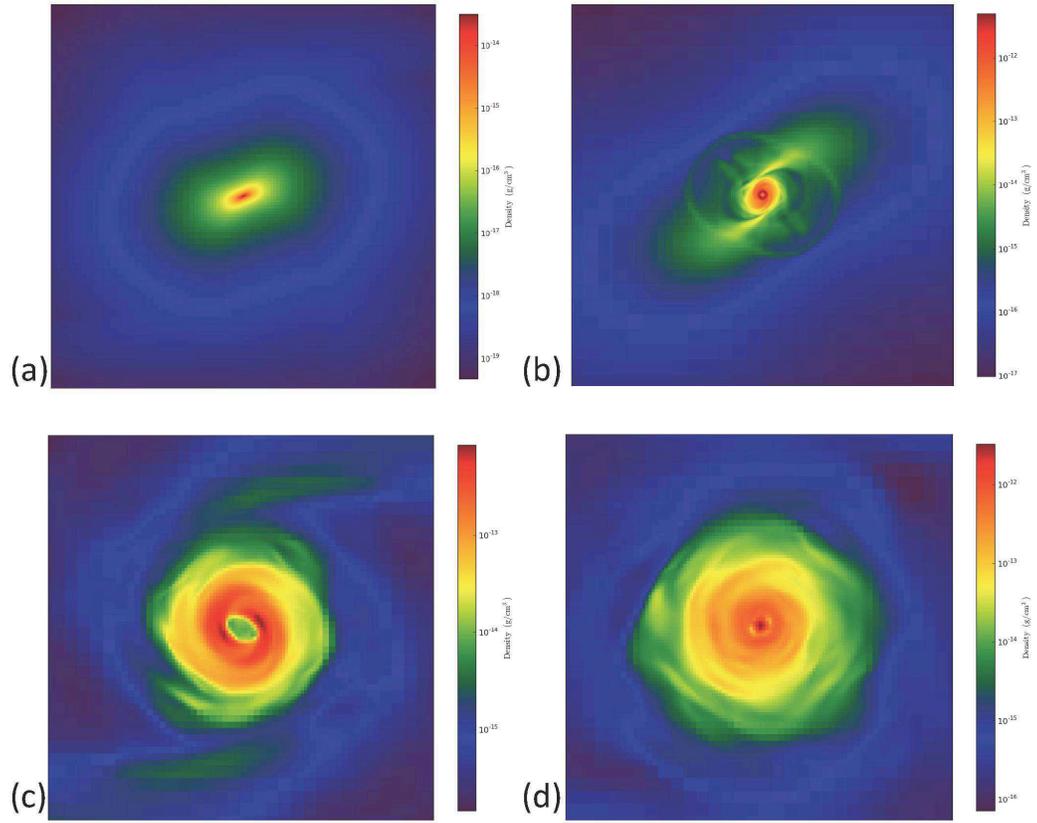}
\caption{Evolution of the midplane ($z = 0$) density distribution for model BMPP2A, shown 
at four times: (a) 1.42 $t_{ff}$, (b) 1.51 $t_{ff}$, (c) 1.56 $t_{ff}$, and (d) 1.62 $t_{ff}$. 
Region shown is $2.0 \times 10^{17}$ cm across in (a), $2.0 \times 10^{16}$ cm across 
in (b), and $6.7 \times 10^{15}$ cm across in (c) and (d).}
\end{figure}

\clearpage

\begin{figure}
\vspace{-2.0in}
\plotone{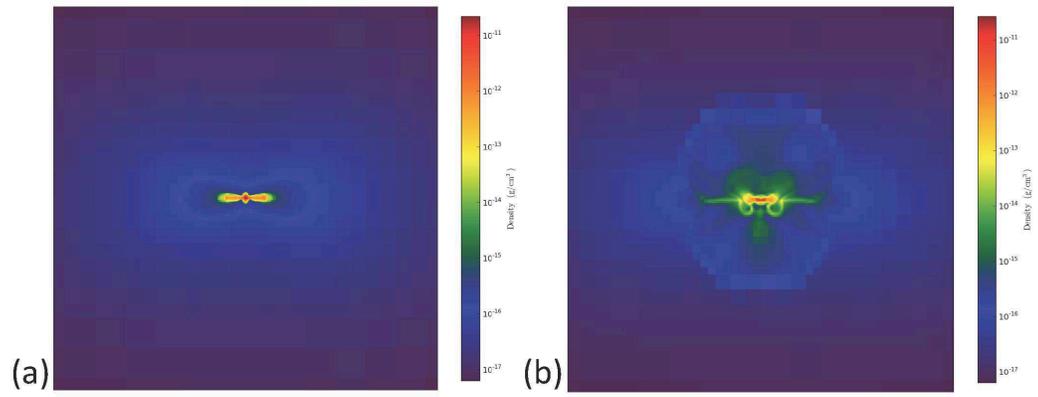}
\caption{Density distributions in the $x = 0$ plane for models (a) PP2A and (b) BMPP2A, 
both shown at a time of 1.51 $t_{ff}$ and both in a region $2.0 \times 10^{16}$ cm across.}
\end{figure}

\clearpage

\begin{figure}
\vspace{-2.0in}
\plotone{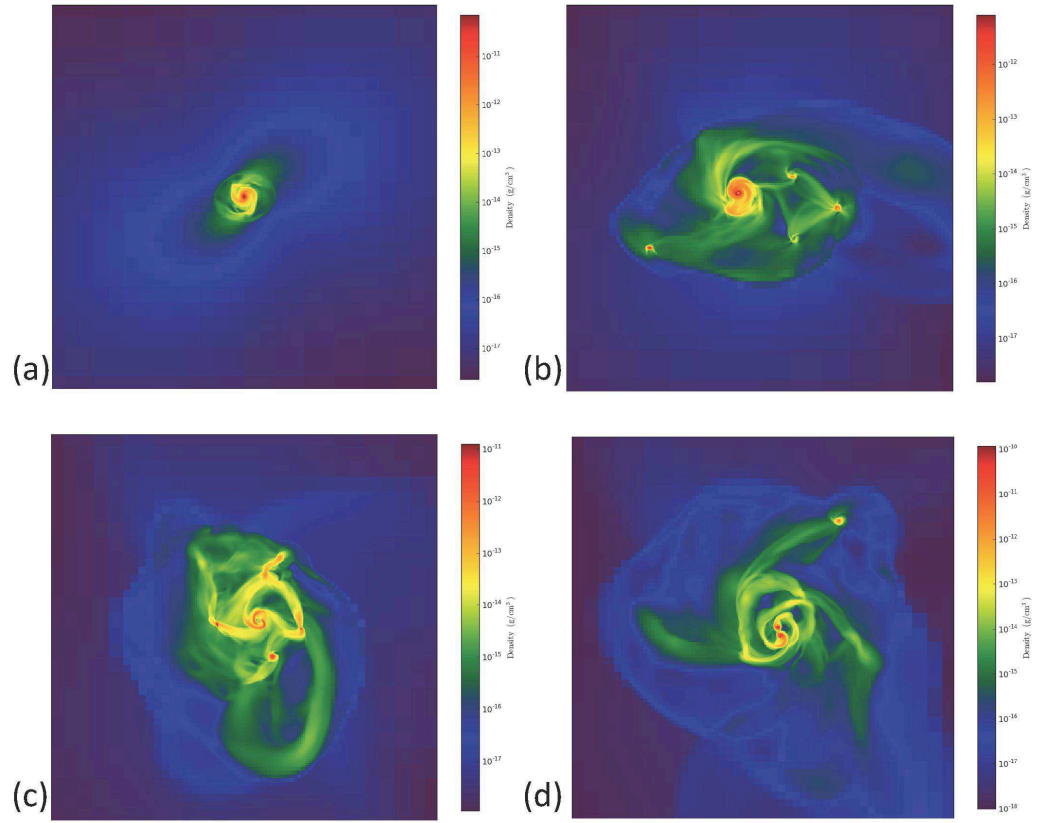}
\caption{Evolution of the midplane ($z = 0$) density distribution for model BNPP2A, shown 
at four times: (a) 1.56 $t_{ff}$, (b) 1.84 $t_{ff}$, (c) 2.05 $t_{ff}$, and (d) 2.17 $t_{ff}$. 
Region shown is $4.0 \times 10^{16}$ cm across at all times.}
\end{figure}

\clearpage

\begin{figure}
\vspace{-2.0in}
\plotone{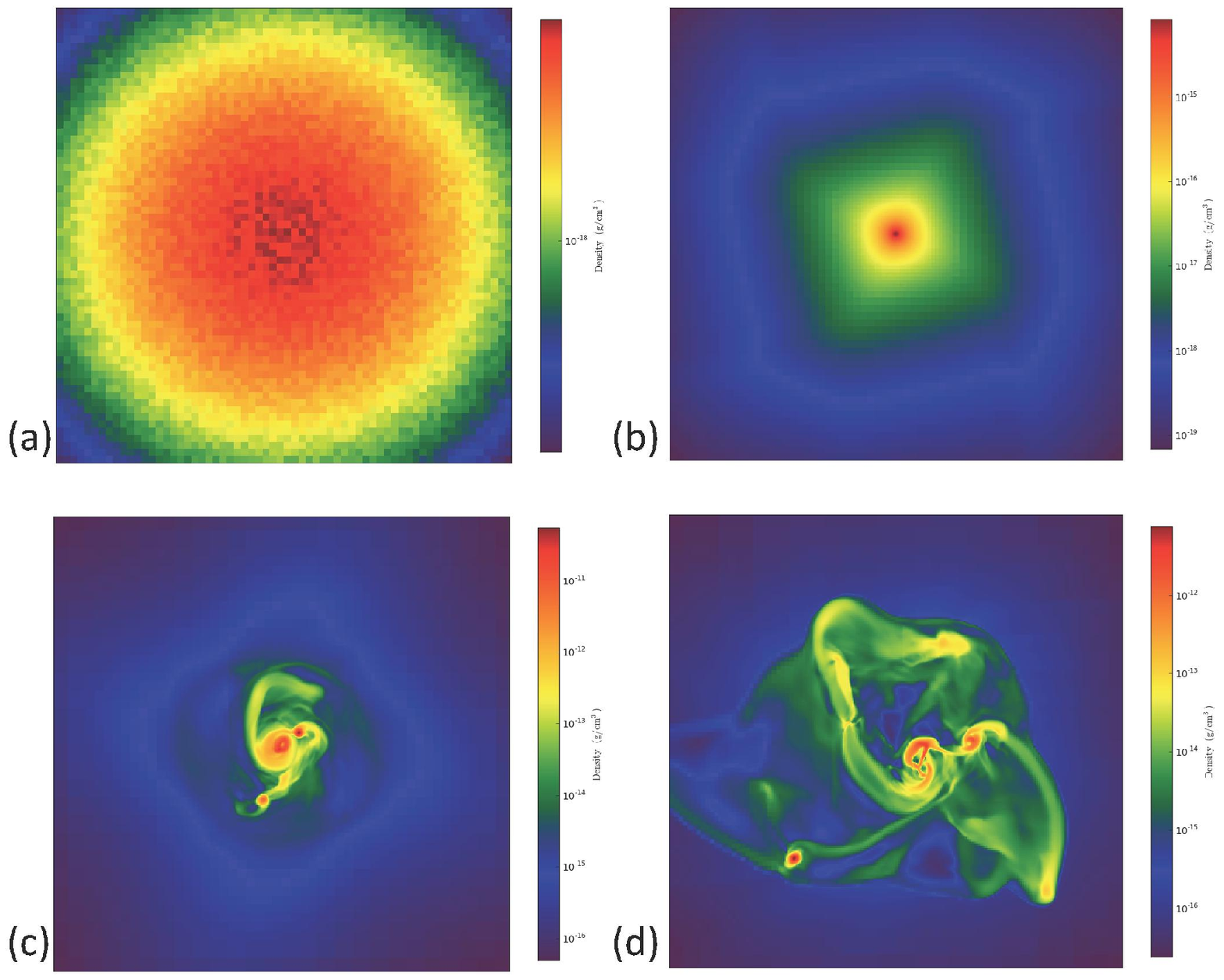}
\caption{Evolution of the midplane ($z = 0$) density distribution for model BNPO2A, shown 
at four times: (a) 0.0, (b) 1.21 $t_{ff}$, (c) 1.36 $t_{ff}$, and (d) 1.51 $t_{ff}$. 
Region shown is $2.0 \times 10^{17}$ cm across in (a) and (b) and 
$2.0 \times 10^{16}$ cm across in (c) and (d).}
\end{figure}

\clearpage

\begin{figure}
\vspace{-2.0in}
\plotone{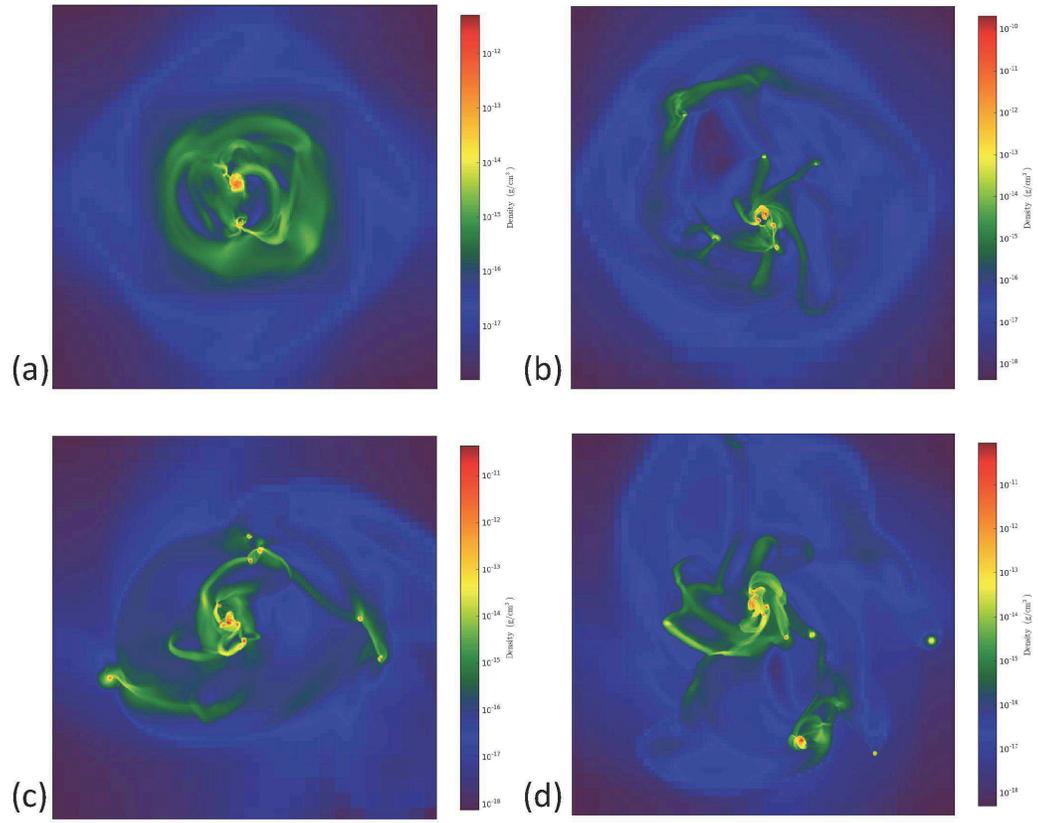}
\caption{Density distributions in the $z = 0$ plane for models (a) BMPO2R at 2.03 $t_{ff}$,
(b) BNPO2R at 2.38 $t_{ff}$, (c) BOPO2R at 2.15 $t_{ff}$, and (d) PO2R at 2.31 $t_{ff}$,
all shown in a region $8.0 \times 10^{16}$ cm across.}
\end{figure}

\clearpage

\begin{figure}
\vspace{-2.0in}
\plotone{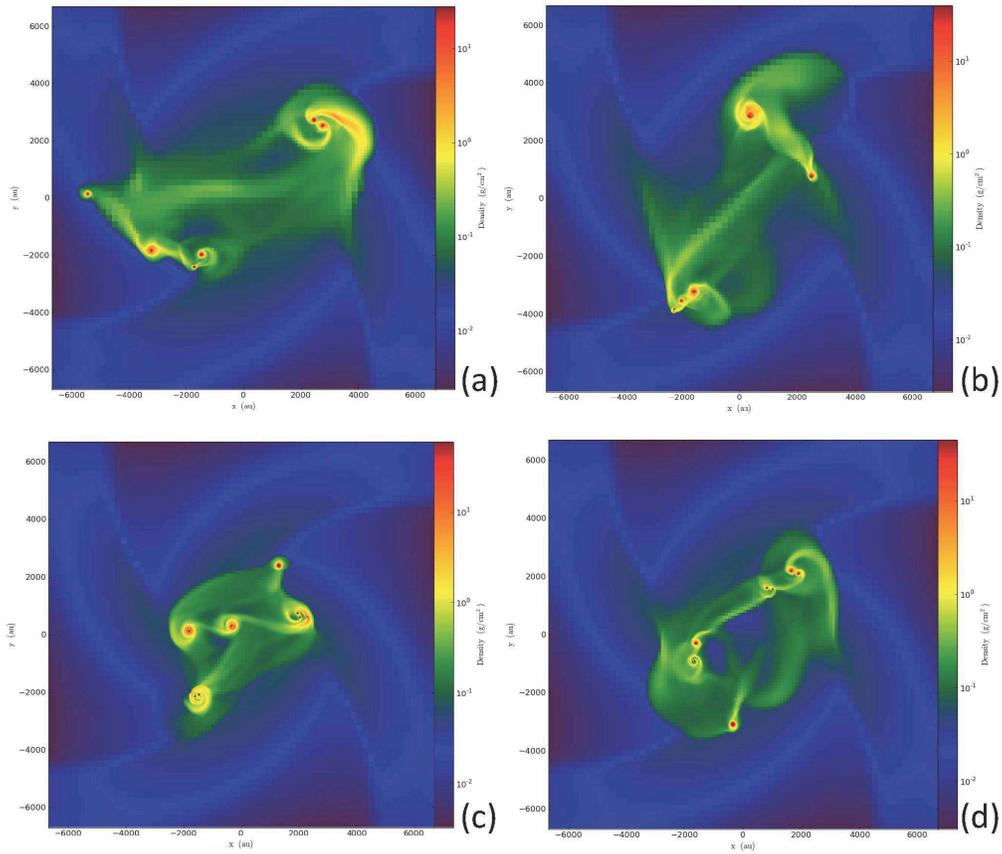}
\caption{Column density distributions (in g cm$^{-2}$) and sink particles (black dots) 
projected onto the $z = 0$ plane for models (a) BOPO2RSP at 2.69 $t_{ff}$,
(b) BOPO2RSQ at 2.61 $t_{ff}$, (c) BOPO2RSR at 2.54 $t_{ff}$, and (d) BOPO2RSS at 2.57 $t_{ff}$,
all shown in a region $2.0 \times 10^{17}$ cm across.}
\end{figure}

\clearpage

\begin{figure}
\vspace{-2.0in}
\plotone{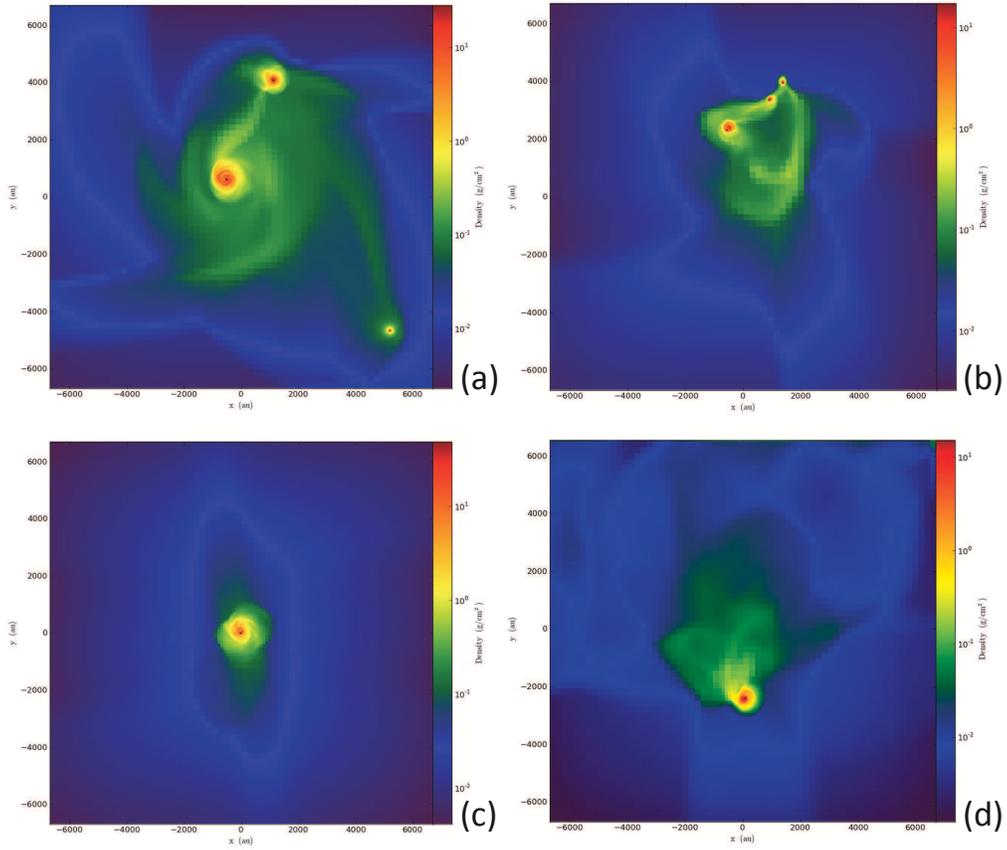}
\caption{Column density distributions (in g cm$^{-2}$) and sink particles (black dots) 
projected onto the $z = 0$ plane for models (a) BOPP2RS at 4.30 $t_{ff}$,
(b) BOPP2AS at 3.11 $t_{ff}$, (c) BOPP2CS at 2.60 $t_{ff}$, and (d) BOPP2ES at 3.63 $t_{ff}$,
all shown in a region $2.0 \times 10^{17}$ cm across.}
\end{figure}

\clearpage

\begin{figure}
\vspace{-2.0in}
\plotone{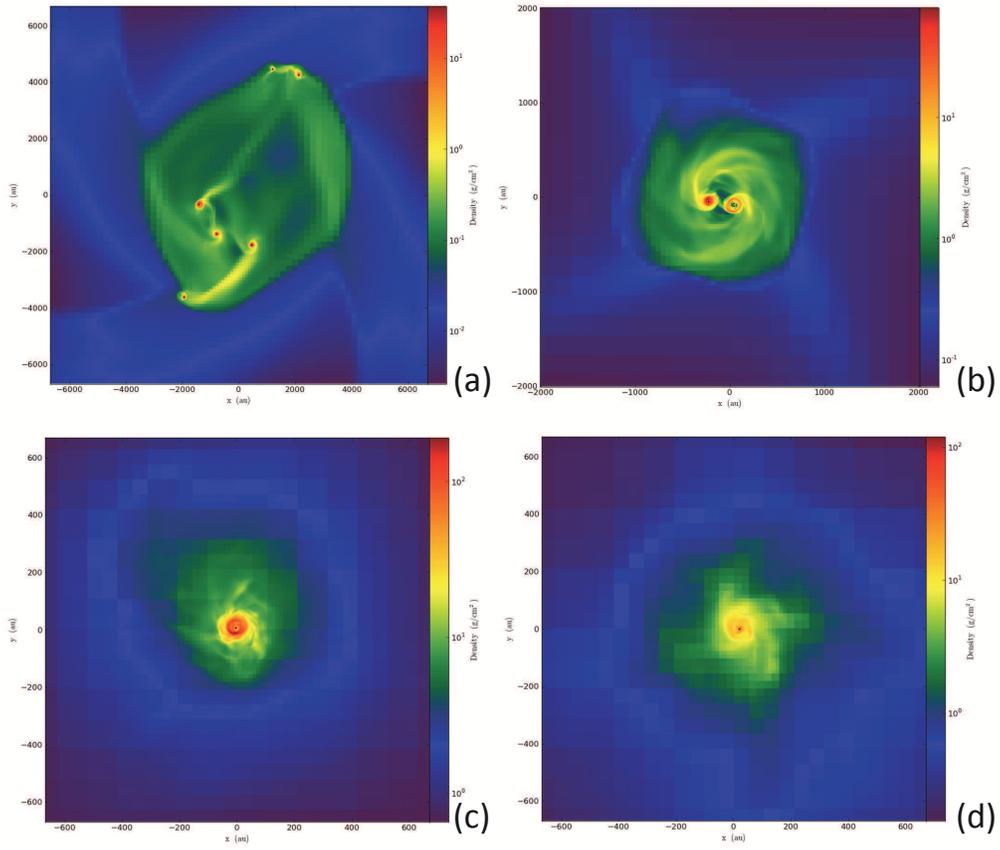}
\caption{Column density distributions (in g cm$^{-2}$) and sink particles (black dots) 
projected onto the $z = 0$ plane for models (a) BNPO2RS at 2.72 $t_{ff}$,
(b) BNPO2AS at 1.84 $t_{ff}$, (c) BNPO2CS at 1.47 $t_{ff}$, and (d) BNPO2ES at 1.85 $t_{ff}$,
Region shown is $2.0 \times 10^{17}$ cm across in (a), $6.0 \times 10^{16}$ cm across 
in (b), and $2.0 \times 10^{16}$ cm across in (c) and (d).}
\end{figure}

\clearpage

\begin{figure}
\vspace{-2.0in}
\plotone{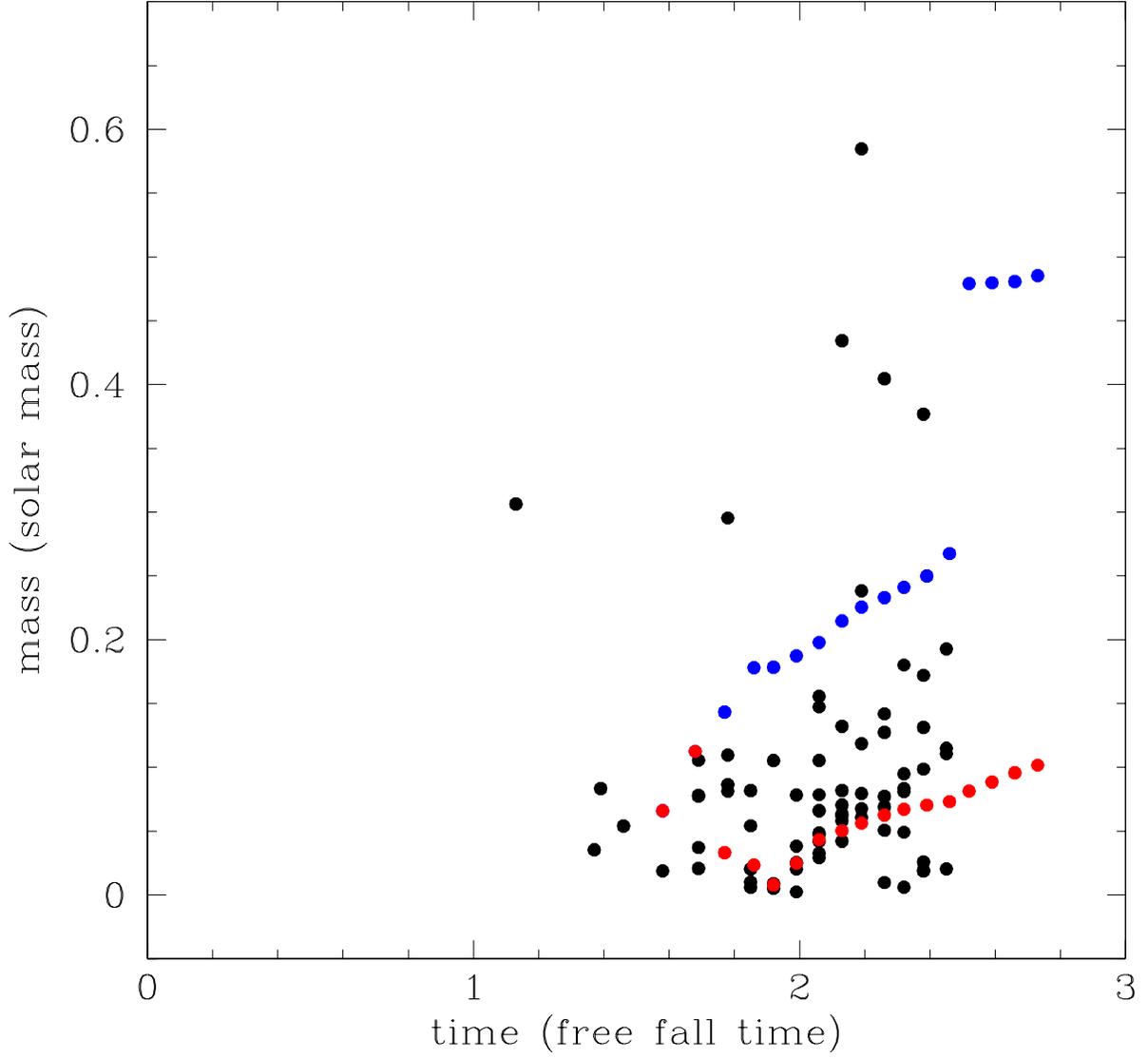}
\caption{Evolution of the clump masses in model BNPO2R (black dots) at various times
compared to the maximum (blue dots) and minimum (red dots) sink particle masses
for model BNPO2RS.}
\end{figure}

\end{document}